\documentclass[3p,times,twocolumn]{elsarticle}

\usepackage{ecrc}

\volume{00}

\firstpage{1}

\journalname{Nuclear Physics B Proceedings Supplement}

\runauth{M.~V.~Carlucci}

\jid{nuphbp}

\jnltitlelogo{Nuclear Physics B Proceedings Supplement}

\usepackage{amssymb}

\usepackage[figuresright]{rotating}

\usepackage{graphicx} 
\usepackage{psfrag} 
\usepackage{subfigure} 
\usepackage{array} 
\usepackage{booktabs} 
\usepackage{amsfonts} 
\usepackage{amssymb} 
\usepackage{amsmath} 
\usepackage{latexsym} 
\usepackage{amsthm} 
\usepackage{slashed} 
\usepackage{bm} 
\usepackage{mathrsfs} 
\usepackage[usenames,dvipsnames]{color} 
\usepackage{subfigure}
\usepackage{color}
\usepackage{paralist}

\newcommand{\be}{\begin{equation}}
\newcommand{\ee}{\end{equation}}

\begin{document}

\begin{frontmatter}



\dochead{}

\title{New Physics scenarios in light of new and old flavour data}


\author{Maria Valentina Carlucci}

\address{Physik-Department, Technische Universit\"at M\"unchen, James-Franck-Stra{\ss}e, D-85748 Garching, Germany}

\begin{abstract}
The flavour observables currently in the spotlight because of the recent experimental updates or because of the presence of tensions with the Standard Model predictions are reviewed in their main aspects. These quantities suggest the development of a particular strategy for testing the viability of Beyond Standard Model scenarios that is applied for a qualitative analysis of different patterns of flavour violation.
\end{abstract}


\end{frontmatter}



\section{Introduction}
The community of flavour physics is currently living a very exciting period because during the last months both the first results coming from the LHC experiments and the last analysis from Tevatron and SLAC have been released. Certainly, the greatest expectations were set on the outcomes of the new collider, but they have been in some sense disappointed, because already from the beginning LHCb and CMS have controverted some pre-existing New Physics (NP) hints; on the other hand, the American old guards have proved themselves to be still able to reserve surprises. Nevertheless, even if for the moment the new data have not provided any striking evidence of NP, they have confirmed and in some cases strengthened some tensions already pointed out during the last years; being of the level of 1-3$\sigma$, they are afflicting more and more seriously the general picture of the flavour observables. In Section 2 of this contribution we present a review of the most investigated quantities, describing their essential theoretical aspects and the present phenomenological and experimental status.

From this summary a non trivial flavour pattern arises, and decoupling possible NP signatures from experimental fluctuations requires very involved analyses. However, in Section 3 we present a strategy of investigation that permits to perform a qualitative or even semi-quantitative test on different NP models even before complete numerical studies. We apply this scheme in Section 4, where we classify NP models according to their pattern of flavour violation since they can present common peculiar signatures. In this sense, the concept of Minimal Flavour Violation (MFV), with its different formulations and extensive analyses, represents a fundamental theoretical framework.

Despite the amount of new information, in some cases the hadronic uncertainties and the experimental errors do not permit to draw definitive statements about the viability of many NP scenarios yet. An improvement of lattice inputs and the measurements of a certain number of key observables are expected for the next years in order to clarify the landscape of the models beyond the SM \cite{Bediaga:2012py}.

\section{New measures and old tensions in flavour data}

\subsection{New measures}
\begin{itemize}
\item {\bf The mixing-induced CP-violation in the decay $B^0_{s} \rightarrow J/\psi \; \phi$.} 
This decay is considered the golden-plated mode for the measurement of the phase in the $B_s - \bar{B}_s$ mixing, both because of its clean experimental signature, and because of its theoretical characteristics. In fact, for the $B_{q}$ ($q=d,s$) decays that satisfy the conditions that
\begin{inparaenum}
\item[(i)] the final state $f$ is accessible to both $B_q$ and $\bar{B}_q$ mesons,
\item[(ii)] $f$ is a CP eigenstate, and
\item[(iii)] the decay is dominated by $b \rightarrow c \bar{c} s$ tree amplitude,
\end{inparaenum}
it can be shown that the direct CP violation vanishes, and that from the measurement of the time-dependent asymmetry one can extract the mixing-induced CP asymmetry $S_f = \sin \phi_q$ \cite{Dunietz:2000cr}. In addition, in these particular conditions, in the SM the phases $\phi_q$ give directly the unitarity triangle angles, $\phi_d \approx 2 \beta$ and $\phi_s \approx - 2 \beta_s$. The decay $B^0_{s} \rightarrow J/\psi \; \phi$ presents all the required characteristics, after that an angular analysis is performed in order to separate the CP parities of the final states due to their relative angular momentum.

The quantity $S_{\psi \phi}$ has been under great attention in the last years, since in 2007 the measurements by CDF and D0 presented a discrepancy larger than 3$\sigma$ with respect to the SM prediction \cite{Bona:2008jn}. In the end of 2011 LHCb presented its first tagged analysis of $B^0_{s} \rightarrow J/\psi \; \phi$ \cite{LHCb:2011aa}; the last results, obtained by the full 2011 data sample of 1.0 fb$^{-1}$ in $pp$ collisions $\sqrt{s} = 7$ TeV, give \cite{LHCb2012}
\begin{subequations}
\be
\phi_s = -0.001 \pm 0.101 \text{(stat)} \pm 0.027 \text{(syst)} \; \text{rad} ~,
\ee
\be
\Delta \Gamma_s = 0.116 \pm 0.018 \text{(stat)} \pm 0.006 \text{(syst)} \; \text{ps}^{-1} ~,
\ee
\end{subequations}
that are the world's most precise measurement of $\phi_s$ and the first direct observation of a non-zero value for $\Delta \Gamma_s$; moreover, the sign of $\Delta \Gamma_s$ has been determined for the first time, found to be positive at 4.7$\sigma$ confidence level \cite{Aaij:2012eq}. These values are fully compatible with the SM predictions, whose updated values read \cite{Lenz:2011ti}
\begin{subequations}
\be
\phi_s = 0.0038 \pm 0.0010 \; \text{rad} ~, 
\ee
\be
\Delta \Gamma_s = 0.087 \pm 0.021 \; \text{ps}^{-1} ~.
\ee
\end{subequations}
\item {\bf The decays $B^0_{(s)} \rightarrow \mu^+ \mu^-$.}
These decays are of particular interest among the electroweak penguin processes, because they are chirality-suppressed in the SM and are most sensitive to scalar and pseudoscalar operators, i.e.~particularly sensitive to the exchange of new (pseudo)scalar particles. In the SM branching ratio the main source of uncertainty is constituted by the $B_s^0$ decay constant $f_{B_s}$, but there has been significant progress in theoretical calculations of this quantity in recent years \cite{Laiho:2009eu}. The most recent predictions are \cite{Buras:2012ru}
\begin{subequations}
\be
\mathcal{B} (B^0_s \rightarrow \mu^+ \mu^-)_{\text{SM}} = (3.23 \pm 0.27) \times 10^{-9} ~,
\ee
\be
\mathcal{B} (B^0_d \rightarrow \mu^+ \mu^-)_{\text{SM}} = (1.07 \pm 0.10) \times 10^{-10} ~;
\ee
\end{subequations}
in using these results a correction has to be applied to $\mathcal{B} (B^0_s \rightarrow \mu^+ \mu^-)_{\text{SM}}$, since it has been recently shown that $\Delta \Gamma_s$ affects the extraction of the SM branching ratio, which need to be rescaled by a factor $r( \Delta \Gamma_s) = 0.91 \pm 0.01$ \cite{DeBruyn:2012wk}.

Until last year, only high upper limits were available for both decays, leaving large space to many NP models, especially the ones with extended Higgs sector, that predict enhanced branching fractions. In 2011 CDF, CMS and LHCb published their new results, and at the beginning of 2012 LHCb set the world best limits \cite{Aaij:2012ac}: at 95\% C.L.,
\begin{subequations}
\be
\mathcal{B} (B^0_s \rightarrow \mu^+ \mu^-) < 4.5 \times 10^{-9} ~,
\ee
\be
\mathcal{B} (B^0_d \rightarrow \mu^+ \mu^-) < 1.0 \times 10^{-9} ~;
\ee
\end{subequations}
that are now very close to the SM predictions.
\end{itemize}

\subsection{Old tensions}
\begin{itemize}
\item {\bf The $\epsilon_K - S_{\psi K_S}$ tension.}
For the reasons explained in the previous paragraph, this decay is the golden mode for the determination of the phase in the $B_s - \bar{B}_d$ mixing, and hence of $\beta$ angle of the unitarity triangle; in fact, in the SM the mixing-induced CP asymmetry reads simply
\be
S_{\psi K_S} = \sin 2 \beta~.
\ee
On the other hand, information about the $\beta$ angle can also be obtained from  the $K-\bar{K}$ system. The relevant CP-violating observable in this case is $\epsilon_K$, and in the SM, considering only the leading top exchange, its module can be written as \cite{Buras:2008nn}
\begin{multline}
|\epsilon_K| = \kappa_\epsilon \frac{G_F^2 M_W^2 M_K F_K^2 \hat{B}_K}{12 \sqrt{2} \pi^2 \Delta M_K} \left( \frac{M_{B_s}}{M_{B_d}} \right) \left( \frac{\Delta M_d}{\Delta M_s} \right) \times \\ \times |V_{cb}|^4 \xi_s^2 \eta_{tt} S_0 (x_t) \sin 2 \beta ~,
\end{multline}
where $k_\epsilon$ accounts for the long-distance effects, $F_K$, $B_K$ and $\xi_s$ are lattice parameters, $\eta_{tt}$ describes the QCD short-distance effects, and $S_0 (x_t)$ is the Inami-Lim loop function. Hence, it is evident how this parameter provides a reasonably clean estimation of $\beta$, since it depends only on experimental quantities, calculable parameters and the CKM element $|V_{cb}|$, that is determined from tree-level processes with small uncertainties.

Already in 2008, motivated by the release of updated values of some lattice parameters, a 2.1$\sigma$ discrepancy between the two determinations of $\beta$ was pointed out \cite{Lunghi:2008aa}. Calculations with higher order terms and penguin contributions have been performed \cite{Brod:2010mj}, and the most recent estimations give \cite{Lunghi:2010gv}
\begin{subequations}
\be
(\sin 2 \beta)_{S_{\psi K_s}} = 0.668\pm 0.023 ~, 
\ee
\be
(\sin 2 \beta)_{\epsilon_K} = 0.867 \pm 0.048 ~.
\ee
\end{subequations}
\item {\bf The determination of $|V_{ub}|$.}
Between the CKM elements, $|V_{ub}|$ is one of the most problematic, since its inclusive and exclusive determinations differ significantly between each other \cite{Beringer:1900zz}:
\begin{subequations}
\be
|V_{ub}| = (4.41 \pm 0.15 \,^{+0.15}_{-0.17}) \times 10^{-3} \quad \text{(inclusive)} ~,
\ee
\be
|V_{ub}| = (3.23 \pm 0.31) \times 10^{-3} \quad \text{(exclusive)} ~.
\ee
\end{subequations}
The inclusive determination of $|V_{ub}|$ comes from $B \rightarrow X_u \ell \bar{\nu}$, and is complicated due to large $B \rightarrow X_c \ell \bar{\nu}$ background. On the other hand, to extract $|V_{ub}|$ from an exclusive channel, like $B \rightarrow \pi \ell \bar{\nu}$, the form factors have to be known. The two determinations are independent.
\item {\bf The branching ratio of $B \rightarrow \tau \; \nu$.}
In the SM, $B \rightarrow \tau \; \nu$ is a simple tree-level decay, and its branching ratio is
\be
\mathcal{B} (B \rightarrow \tau \; \nu) = \frac{G_F^2 m_B m_{\tau}^2}{8 \pi} \left( 1-\frac{m_{\tau}^2}{m_{B}^2} \right)^2 f_B^2 |V_{ub}|^2 \tau_B ~;
\ee
while the Fermi constant, the masses and the $B$ lifetime are precisely measured quantities, we have seen how there are tensions in the determination of $|V_{ub}|$; nevertheless, both its inclusive and its exclusive value lead to discrepancies with its experimental value \cite{Beringer:1900zz}:
\be
\mathcal{B} (B \rightarrow \tau \; \nu)^{\text{exp}} = (1.79 \pm 0.48) \times 10^{-4} ~,
\ee
to be compared with
\begin{subequations}
\be
\mathcal{B} (B \rightarrow \tau \; \nu)^{\text{SM}}_{\text{incl}} = (1.22 \pm 0.31) \times 10^{-4} \quad \text{(1.0}\sigma\text{)} ~,
\ee
\be
\mathcal{B} (B \rightarrow \tau \; \nu)^{\text{SM}}_{\text{excl}} = (0.67 \pm 0.15) \times 10^{-4} \quad \text{(2.9}\sigma\text{)} ~.
\ee
Moreover, if one tries to eliminate the dependency from $|V_{ub}|$ by using the unitarity conditions of the CKM matrix, the branching ratio will depend on the CKM parameters $|V_{ud}|$, $\beta$ and $\gamma$, and the SM prediction will be \cite{Lunghi:2010gv}
\be
\mathcal{B} (B \rightarrow \tau \; \nu)^{\text{SM}}_{\text{fit}} = (0.754 \pm 0.093) \times 10^{-4} ~,
\ee
\end{subequations}
with a 2.8$\sigma$ deviation from the experimental measure.
\item {\bf The anomalous like-sign dimuon charge asymmetry.} The like-sign dimuon charge asymmetry $A^b_{sl}$ for semileptonic decays of $b$ hadrons produced in proton-antiproton collisions is defined as
\be
A^b_{sl} = \frac{N_b^{++}-N_b^{--}}{N_b^{++}+N_b^{--}} ~,
\ee
where $N_b^{++}$ and $N_b^{--}$ are the numbers of events containing two $b$ hadrons that decay semileptonically via $b \rightarrow \mu X$, producing two positive or two negative muons respectively. It can be expressed as \cite{Grossman:2006ce}
\be
A^b_{sl} = \frac{f_d Z_d a^d_{sl}+f_s Z_s a^s_{sl}}{f_d Z_d + f_s Z_s} ~,
\ee
where $Z_q$ are functions of the $B_q$ mixing parameters $\Gamma_q, \Delta M_q$ and $\Delta \Gamma_q$, the quantities $f_q$ are the production fractions for $\bar{b} \rightarrow B_q$, and $a^q_{sl}$ is the charge asymmetry for the ``wrong-charge'' (i.e. a muon charge opposite to the charge of the original $b$ quark) semileptonic $B_q$-meson decay induced by oscillation:
\be
a_{sl}^q = \frac{\Gamma (\bar{B}_q (t) \rightarrow \mu^+ X) - \Gamma (B_q (t) \rightarrow \mu^- X)}{\Gamma (\bar{B}_q (t) \rightarrow \mu^+ X) + \Gamma (B_q (t) \rightarrow \mu^- X)} ~;
\ee
the latter is independent of $t$, and can be written as
\be
a_{sl}^q = \frac{\Delta \Gamma_q}{\Delta M_q} \tan \phi_q ~.
\ee
In this way, substituting the experimental values, the like-sign dimuon charge asymmetry reads \cite{Asner:2010qj}, \cite{Lenz:2012az}
\be
A^b_{sl} = (0.532 \pm 0.039) a_{sl}^d + (0.468 \pm 0.039) a_{sl}^s ~,
\ee
showing more explicitly the dependence on the relevant parameters $\Delta \Gamma_q$ and $\phi_q$.
Using the SM values of $\Delta \Gamma_q$ and $\phi_q$, one obtains \cite{Lenz:2006hd}
\begin{subequations}
\be
a^d_{sl} \text{(SM)} = (-4.8 \, ^{+1.0}_{-1.2}) \times 10^{-4} ~,
\ee
\be
a^s_{sl} \text{(SM)} = (2.1 \pm 0.6) \times 10^{-5} ~,
\ee
\end{subequations}
and the predicted value of $A_{sl}^b$ is
\be
A^b_{sl} \text{(SM)} = (-2.3 \, ^{+0.5}_{-0.6}) \times 10^{-4} ~.
\ee

The most recent determination of the like-sign dimuon asymmetry, obtained in 2010 in 6.1 fb$^{-1}$ of $p\bar{p}$ collisions recorded with the D0 detector at a center-of-mass energy $\sqrt{s} =$ 1.96 TeV \cite{Abazov:2010hv}, gives
\be
A^b_{sl} = -0.00957 \pm 0.00251 \text{(stat)} \pm 0.00146 \text{(syst)} ~,
\ee
and differs by 3.2$\sigma$ from the SM prediction.
\item {\bf  Other tensions.}
Other quantities that present discrepancy between the SM prediction and the experimental measurements that are softer or still subject of work in progress are the following.
\begin{itemize}
\item In the light of the recent lattice inputs \cite{Laiho:2009eu}, both $\Delta M_d$ and $\Delta M_s$ are 1$\sigma$ above the data \cite{Buras:2012ts}.
\item Indications of the SM branching ratio for $B \rightarrow X_s \gamma$ being 1.2$\sigma$ below the data, for $B \rightarrow X_s \ell^+ \ell^-$ at high $q^2$ being visibly below the data too, and for the $K^*$ longitudinal polarization fraction in $B \rightarrow K^* \ell^+ \ell^-$ being larger than data \cite{Buras:2012ts} are currently under careful experimental investigation \cite{:2012iw}, \cite{:2012vwa}.
\item The very recent measurements of the ratios $R(D^{(}{}^*{}^{)}) = \mathcal{B}(B \rightarrow D^{(}{}^*{}^{)} \tau \, \nu)/ \mathcal{B}(B \rightarrow D^{(}{}^*{}^{)} \ell \, \nu)$, where $\ell$ is either $e$ or $\mu$, exceed the SM expectations by 2.0$\sigma$ and 2.7$\sigma$ respectively \cite{Lees:2012xj}.
\end{itemize}
\end{itemize}

\section{Strategy of investigation}
The task of recognizing the pattern of deviations of the present experimental data from the SM expectations is non trivial because of the $\epsilon_K - S_{\psi K_S}$ tension. In fact, the picture is noticeably different depending on whether $\epsilon_K$ or $S_{\psi K_S}$ is used as a basic observable to fit the $\beta$ angle of the CKM matrix; both quantities can receive important NP contributions, and the one in which NP is required depends on the values of the CKM parameters $\gamma$ and $|V_{ub}|$ \cite{Buras:2008nn}, \cite{Buras:2009pj}. Now, the angle $\gamma$ is known with a very low accuracy, $\gamma = (68^{+10}_{-11})^{\circ}$ \cite{Beringer:1900zz}, and discrepancy between the inclusive and exclusive determinations of $|V_{ub}|$ makes the average between the two results a poorly significant value.

Waiting for more accurate values of these tree-level parameters, a possible strategy is to set the angle $\gamma \approx 68^{\circ}$ and to consider separately two possible scenarios:
\begin{description}
\item[Scenario 1]
\end{description}
\begin{itemize}
\item Exclusive (small) $|V_{ub}|$
\item $S_{\psi K_S}$ in agreement with data
\item $\epsilon_K$ below data of $\sim 1-2\sigma$
\item $\mathcal{B} (B \rightarrow \tau \, \nu)$ below data of $\sim 3\sigma$
\end{itemize}
\begin{description}
\item[Scenario 2]
\end{description}
\begin{itemize}
\item Inclusive (large) $|V_{ub}|$
\item $S_{\psi K_S}$ above data of $\sim 2-3\sigma$
\item $\epsilon_K$ in agreement
\item $\mathcal{B} (B \rightarrow \tau \, \nu)$ below data of $\sim 1\sigma$
\end{itemize}

Once this framework has been established, the first step in the phenomenological study of a NP model will be to determine if the experimental constraints can be satisfied in both the $|V_{ub}|$ scenarios or if the model selects a particular $|V_{ub}|$ value. Of course, the simplest SM extensions, having a small number of degrees of freedom, will be put under pressure by this procedure, while more elaborated models will have more chance to survive the test but will be at the same time less predictive. Once one scenario has been chosen, one can study the remaining freedom in the space parameters in order to relax the remaining tensions.

\section{Patterns of flavour violation}
Despite the discrepancies described above, the success of the SM predictions in the flavour sector remains impressive; this tends to push the scale of NP up to $\mathcal{O}(10^2)-\mathcal{O}(10^3)$ TeV, while the stabilization of the SM would require some new mechanism already at the TeV scale. However, even a NP model at the TeV scale could describe correctly and naturally the experimental data if it had the same successful flavour structure as the SM. This idea has lead to define different classes of flavour violation patterns in the NP scenarios according to how much they are similar to the SM flavour pattern, and to identify their common features and peculiar signatures.

\subsection{Constrained Minimal Flavour Violation}
This is the most pragmatic approach to flavour violation, and is satisfied by the simplest SM extensions. It is defined by two conditions \cite{Buras:2000dm}, \cite{Blanke:2006ig}:
\begin{itemize}
\item all flavour changing transitions are governed by the CKM matrix with the CKM phase being the only source of CP violation;
\item the only relevant operators in the effective Hamiltonian below the weak scale are those that are also relevant in the SM.
\end{itemize}
There are basically three main implications of these assumptions: 
\begin{itemize}
\item $S_{\psi K_s}$ and $S_{\psi \phi}$ are as in the SM;
\item For fixed CKM parameters determined in tree-level decays, $|\epsilon_K|$, $\Delta M_d$ and $\Delta M_s$ can only be enhanced relative to SM predictions, and this happens in a correlated manner \cite{Blanke:2006yh}.
\item There are correlations between various observables \cite{Buras:2003jf}; for example, in this context the most relevant are
\begin{subequations}
\be
\frac{\Delta M_d}{\Delta M_s} = \frac{M_{B_d}}{M_{B_s}} \frac{\hat{B}_d}{\hat{B}_s} \frac{F^2_{B_s}}{F^2_{B_s}} \left| \frac{V_{td}}{V_{ts}} \right|^2 r(\Delta M) ~,
\ee
\be
\frac{\mathcal{B} (B^0_s \rightarrow \mu^+ \mu^-)}{\mathcal{B} (B^0 \rightarrow \mu^+ \mu^-)} = \frac{\tau(B_s)}{\tau(B_d)} \frac{M_{B_s}}{M_{Bs}} \frac{F^2_{B_d}}{F^2_{B_s}} \left| \frac{V_{td}}{V_{ts}} \right|^2 r(\mu^+ \mu^-)
\ee
\end{subequations}
where $r(\Delta M)=r(\mu^+ \mu^-)=1$ in CMFV and deviations from unity can be used to recognize and parametrize different patterns of flavour violation. In this sense these relations can be regarded as \emph{standard candles of flavour physics}.
\end{itemize}

Following the analysis strategy outlined in the previous paragraph, we start from the observation that since $S_{\psi K_S}$ cannot receive new contributions in this class of models, constrained MFV prefers the Scenario 1 for $|V_{ub}|$. On the other hand, we have seen that it allows the enhancement of $\epsilon_K$, and hence the $S_{\psi K_S}-\epsilon_K$ tension can be solved. Nevertheless, it has been shown that the enhancement of $\epsilon_K$ would determine a correlated enhancement of both $\Delta M_d$ and $\Delta M_s$ \cite{Buras:2012ts}, which are already slightly above the experimental values.

The previous considerations, even if only qualitative, point out the difficulties that constrained MFV models have in accommodating the tensions in flavour data, due to the presence of few free parameters and strict correlations. More quantitative studies, as well as a complete analysis of more observables, could be already able to derive more definitive statements about the viability of this flavour violation scheme.

\subsection{Minimal Flavour Violation at large}
With a more formal approach to the flavour violation issue, one recognizes that the gauge part of the SM quark Lagrangian presents a large global flavour symmetry $\mathcal{G}_q = (SU(3) \times U(1))^3$ (i.e.~ a family $SU(3)$ and a phase for each electroweak multiplet), and that this symmetry is explicitly broken in the Higgs sector only by the Yukawa couplings $Y_u$ and $Y_d$. This is the specific successful symmetry plus symmetry breaking pattern of the SM, and in order to recognize it in a NP model one can promote the Yukawa couplings to \emph{spurions} $Y_u \sim (3,\bar{3},1)_{SU(3)^3}$, $Y_d \sim (3,1\bar{3})_{SU(3)^3}$ that formally recover the SM flavour symmetry; then, one will say that a theory satisfies the MFV criterion if it is formally invariant under the global flavour symmetry $\mathcal{G}_q$ as the SM \cite{D'Ambrosio:2002ex}. The phenomenological implications of MFV can be drown out using an effective field theory approach, that is building higher dimensional operators with spurions and studying their parameters which contain the NP effects. Since these parameters can be complex in general, \emph{flavour-blind} CP violating new phases can be present.

An example of application of this approach is the 2HDM$_{\overline{\text{MFV}}}$. In a general 2-Higgs-duoblet model, the new neutral Higgs fields can mediate large FCNCs; if instead MFV is imposed, the effective down-type FCNC interaction term is
\be
{\bar d}^i_L \left[
\left( a_0 V^\dagger \lambda_u^2 V + a_1 V^\dagger \lambda_u^2 V  \Delta + a_2  \Delta V^\dagger \lambda_u^2 V \right) \lambda_d \right]_{ij}
 d^j_R~H~,
\ee
where $\lambda_{u,d} \propto 1/v \; \text{diag} \left( m_{u,d},m_{c,s},m_{t,b} \right)$, $\Delta = \text{diag} \left( 0,0,1 \right)$, and the $a_i$ are complex parameters of $\mathcal{O} (1)$; because of the presence of two off-diagonal CKM elements and the down-type Yukawas, it has been shown that FCNCs are suppressed in an effective and natural way \cite{Buras:2010mh}. Moreover, the main features of the relevant phenomenological quantities are the following \cite{Buras:2010mh}:
\begin{itemize}
\item the impact in $K$, $B$ and $B_s$ mixing amplitudes scales with $m_s m_d$, $m_b m_d$ and $m_b m_s$ respectively;
\item new flavour-blind phases can contribute to the $B$ and $B_s$ systems in the following way:
\be
S_{\psi K_S} = \sin (2 \beta + 2 \phi_{B_d}) ~, \quad S_{\psi \phi} = \sin (2 |\beta_s| - 2 \phi_{B_s}) ~,
\ee
with $\phi_{B_s} = (m_s/m_d) \phi_{B_d}$; instead they are not present in the $K$ system.
\end{itemize}

The previous observations imply that $\epsilon_K$ can receive only tiny new contributions while $S_{\psi K_S}$ could be in principle sizably modified, and hence the 2HDM$_{\overline{\text{MFV}}}$ selects the Scenario 2 for $|V_{ub}|$. However, a suppression of $S_{\psi K_S}$ would determine a correlated enhancement of $S_{\psi \phi}$, a consequence that was considered very welcome until last year when LHCb put an end to the hopes of new physics in the $B_s$ mixing phase and that therefore puts this model in difficulty. 

Flavour-blind phases can be present in the two-Higgs potential too, giving $\phi_{B_s} = \phi_{B_d}$ \cite{Buras:2010zm}, and could be used to remove the $S_{\psi K_S} - \epsilon_K$ anomaly, but the size of $\phi_{B_d}$ that is necessary would imply in turn $S_{\psi \phi} > 0.15$, which is 2$\sigma$ away from the LHCb central value \cite{Buras:2012ts}.

\subsection{Beyond Minimal Flavour Violation}
Models that do not satisfy the MFV hypotheses, neither by construction nor by imposition, present in general large flavour violating effect, unless fine tuning is applied or some other suppression mechanism occurs.

The latter is the case of a particular class of models with Gauged Flavour Symmetries \cite{Grinstein:2010ve}. The appealing idea at the base of this scenarios is the assumption that the SM flavour symmetry $\mathcal{G}_q$ is a true symmetry of Nature, spontaneously broken by two new scalar fields $Y_u$ and $Y_d$ called \emph{flavons}. This symmetry must be a gauge symmetry in order to avoid the presence of Goldstone bosons, and hence new flavour-mediating gauge bosons are present; moreover, 4 new quarks have to be added in order to make the theory anomaly-free. At this point, the nice feature that one finds is that with this minimal particle content, and in particular because of the presence of the new quark fields, a mechanism of \emph{inverted hierarchy} in the masses is automatically at work, being able to effectively suppress the FCNCs generated by the flavour gauge bosons.

Because of the presence of many new particles and complex free parameters, analytic statements and simple correlations cannot be provided as in the previous cases; nevertheless, an accurate numerical comparison of the new effects with the experimental data permits to strongly constrain the parameter space \cite{Buras:2011wi}. In particular, the following aspects have been pointed out.
\begin{itemize}
\item Large corrections to all the CP observables in the meson oscillation, $\epsilon_K$, $S_{\psi K_S}$ and $S_{\psi \phi}$, are allowed; in particular, requiring an enhancement of $\epsilon_K$ in order to agree with data determines only small variations to $S_{\psi K_S}$ and $S_{\psi \phi}$, a result that is in agreement with the last LHCb results for $S_{\psi \phi}$ and that indicates that this model prefers the Scenario 1 for $|V_{ub}|$.
\item The values of $\Delta M_d$ and $\Delta M_s$ are strongly correlated with $\epsilon_K$, and even if still within 3$\sigma$ after the recent lattice updates, the agreement with the experimental values is very problematic.
\item The agreement of the dimuon asymmetry $A^b_{sl}$ can be slightly improved, while the discrepancy of $\mathcal{B} (B \rightarrow \tau \, \nu)$ is even worsened.
\end{itemize}
In summary, even if the new LHCb data and the recent lattice inputs provide a relief for this scenario, there are still many flavour observables that put it under strong pressure.

\section*{Acnowledgments}
I would like to thank Andrzej Buras for giving me the possibility to contribute to this Workshop, and Giulia Ricciardi for the great organization work and her kind availability. This work has been supported in part by the Graduiertenkolleg GRK 1054 of DFG.







\end{document}